\documentclass{article}
\usepackage{amsfonts}
\usepackage{amssymb}
\usepackage{amsmath}
\usepackage{amsthm}
\usepackage[english]{babel}
\begin{document}

\title{\bf On the Hamiltonian analysis\\ of non-linear massive gravity}

\author{{\bf Alexey Golovnev}\\
{\small {\it Saint-Petersburg State University, high energy physics department,}}\\
{\small \it Ulyanovskaya ul., d. 1; 198504 Saint-Petersburg, Petrodvoretz; Russia}\\
{\small agolovnev@yandex.ru, \quad golovnev@hep.phys.spbu.ru}}
\date{}

\maketitle

\begin{abstract}

In this paper we present a very simple and independent argument for the absence of the Boulware-Deser ghost in
the recently proposed potentially ghost-free non-linear massive gravity. The limitation is that, in its simple form,
the argument is, in a sense, non-constructive and less explicit than the standard approach. However, the 
formalism developed here may prove to be useful for discussing the formal aspects of the theory.

\end{abstract}

\vspace{1cm}

\section{Introduction}

It has been known for the very long time that giving a mass to the graviton in a stable and viable manner
is a very difficult task, if not impossible. At the linear level the only healthy mass term is the one
found by Fierz and Pauli in their classical paper \cite{FP}. It propagates only five degrees of freedom
in accordance with the general expectations for a massive spin-2 particle, while generically there would
be six independent variables in the theory, with the sixth one representing a ghost. Unfortunately, the linear
theory anyway contradicts observations due to the scalar graviton which couples to dust modifying the
effective gravitational constant, but not to radiation keeping the bending of light intact. It was however later argued
by Vainshtein \cite{Vain} that non-linear effects will take over at small scales and save the whole day.
But, almost at the same time, Boulware and Deser have shown \cite{BD} that even with the Fierz-Pauli mass term,
be there any non-linear Vainshtein mechanism or not, the sixth
degree of freedom comes back at the non-linear level reintroducing the ghost mode. And therefore a
stable theory of massive gravity is probably not possible at all.

However, recently a proposal for a ghost-free massive gravity has appeared \cite{dRG,dRGT,HR}. This theory
has been extensively analysed in the perturbation theory, and the absence of the sixth mode
was shown explicitly up to the fourth order in perturbations \cite{dRGT2,dRGT3}. At the same time,
a fully non-linear Hamiltonian analysis has been done and proved the existence of the Hamiltonian
constraint non-perturbatively \cite{HR2,HR3}. After that, some doubts were expressed in the literature
\cite{Kluson} as to the existence of the secondary constraint generated by the Hamiltonian one which
is needed for a consistent elimination of the sixth degree of freedom. But very recently
the secondary constraint in non-linear massive gravity was evaluated (almost) explicitly \cite{HR4} thus
finalising the proof of absence of the Boulware-Deser ghost in this class of models. It justifies the large
amont of interest which has been drawn towards understanding the phenomenological consequences
of the new model. And now we have a rich literature discussing the black holes \cite{KNT,TNieu,KNT2,BHdRGT} and cosmological
solutions \cite{CosdRGT,Lin,obsHassan} in the massive gravity, as well as the first interesting results
concerning the cosmological perturbations \cite{Lin2}.

The ADM analysis of \cite{HR2,HR3,HR4} is technically quite involved; and given the paramount
importance of the topic, we find it necessary to have a thorough understanding of this
non-linear phenomenon of ghost exorcision from various vantage points. In this paper we would like
to offer very simple arguments for the absence of the sixth degree of freedom in de Rham-Gabadadze-Tolley 
(dRGT) gravity at the fully non-perturbative level. In Section 2 we give a brief review of the standard
ADM analysis and its application to massive gravity models. In Section 3 we introduce our set-up, and
explain a very simple reason for the Hamiltonian constraint to exist, at least in the minimal
dRGT-gravity with the flat reference metric. In Section 4 we comment on the general dRGT models. And finally, in Section 5
we conclude.

\section{Review of the ADM Hamiltonian analysis}

The standard way \cite{ADM} of doing the Hamiltonian analysis in GR is via the (3+1)-decomposition of
space-time:
\begin{equation}
\label{ADM}
ds^2\equiv g_{\mu\nu}dx^{\mu}dx{\nu}=-(N^2-N_k N^k)dt^2+2N_i dx^i dt+\gamma_{ij}dx^i dx^j
\end{equation}
where $N$ and $N_i$ are the lapse and shift functions respectively, and $N_i\equiv\gamma_{ik}N^k$.
One can also find the inverse metric in terms of the lapse and shift functions and
inverse spatial metric:
\begin{eqnarray}
\label{ADMinv}
g^{\mu\nu} = 
\left( \begin{array}{cc}
-\frac{1}{N^2} & \frac{N^i}{N^2} \\
\frac{N^j}{N^2} & \gamma^{ij}-\frac{N^i N^j}{N^2} 
\end{array} \right).
\end{eqnarray}
And as the $M_{00}$-minor of the $g_{\mu\nu}$ matrix equals $\gamma\equiv{\rm det}\gamma_{ij}$, we
can conclude from $g^{00}=-\frac{1}{N^2}$ that $\sqrt{-g}=N\sqrt{\gamma}$ due to the standard rule
of inverting the matrices. 

The next step is to calculate the Einstein-Hilbert Lagrangian density in terms
of the ADM variables which is not so easy a task unless one takes the general geometric relations for embedded geometries
directly from the textbooks.
We find it most reasonable to recall the definition of the Riemann tensor as a commutator of
covariant derivatives, and then to follow the geometric path outlined in the classical
volume by Misner, Thorne and Wheeler \cite{MTW}.
But anyway, the result is
\begin{equation*}
\sqrt{-g}R=N\sqrt{\gamma}\left(\mathop{R}\limits^{({\mathit 3})}+K^i_k K^k_i - (K^i_i)^2\right)
+({\rm total\ time\ derivative\ and\ covariant\ divergence\ terms})
\end{equation*}
where we have introduced the extrinsic curvatures
\begin{equation}
\label{ext}
K_{ik}=\frac{1}{2N}\left(\mathop{{\bigtriangledown}_i}\limits^{({\mathit 3})}N_k 
+\mathop{{\bigtriangledown}_k}\limits^{({\mathit 3})}N_i-\dot{\gamma}_{ik}\right),
\end{equation}
and the three-dimensional scalar curvature and covariant derivatives as well as raising and lowering of the indices
are defined by the spatial slice metric $\gamma_{ij}$.

Now one can define the canonical momenta $\pi^{ij}\equiv\frac{\partial {\mathit L}}{\partial \dot{\gamma_{ij}}}$ for the physical variables
and also find the primary constraints $\pi_N=0$ and $\pi_{N_i}=0$ for the momenta of the lapse and shift functions which act
as Lagrange multipliers. The Hamiltonian then reads
\begin{equation}
\label{ham}
H=-\int d^3 x\sqrt{\gamma}\left(N\left(\mathop{R}\limits^{({\mathit 3})}+\frac{1}{\gamma}\left(\frac12 \left(\pi^j_j\right)^2-\pi_{ik}\pi^{ik}\right)\right)
+2N^i\mathop{{\bigtriangledown}^k}\limits^{({\mathit 3})}\pi_{ik} \right).
\end{equation}
The commutation of this Hamiltonian with the unphysical momenta directly gives the four physical constraints,
${\mathcal C}=-\sqrt{\gamma}\left(\mathop{R}\limits^{({\mathit 3})}+\frac{1}{\gamma}\left(\frac12 \left(\pi^j_j\right)^2
-\pi_{ik}\pi^{ik}\right)\right)$ and 
${\mathcal C}_i=-2\sqrt{\gamma}\mathop{{\bigtriangledown}^k}\limits^{({\mathit 3})}\pi_{ik}$, 
which reduce the number of propagating degrees of freedom from six to two, and make the Hamiltonian
equal zero (in the weak sense of Dirac) as it should be in a time-reparametrization-invariant theory.

\subsection{dRGT gravity}

In the massive gravity theories one has to introduce an additional reference metric $f_{\mu\nu}$ which can either be taken fixed
(for example, Minkowski one) or endowed with its own dynamics thus producing a bigravity
model, for otherwise there is no way to construct a non-derivative invariant which would act as
a potential term for the graviton. (We will exclusively take the fixed $f_{\mu\nu}$ metric option.) 
The basic ingredient of the dRGT model is the square-root
matrix $\left(\sqrt{g^{-1}f}\right)^{\mu}_{\nu}$, and the minimal model potential is taken to be \cite{HR}
\begin{equation}
\label{potential}
V=2m^2\left(\left(\sqrt{g^{-1}f}\right)^{\mu}_{\mu}-3\right).
\end{equation}
In what follows we will ignore the $-6m^2$ term which is introduced in order to avoid a contribution
to the cosmological constant. And in the simplest case of Minkowski reference metric we deal with
the square root of the following matrix:
\begin{eqnarray}
\label{basic}
g^{\mu\alpha}\eta_{\alpha\nu}=
\left( \begin{array}{cc}
\frac{1}{N^2} & \frac{N^i}{N^2} \\
-\frac{N^j}{N^2} & \gamma^{ij}-\frac{N^i N^j}{N^2} 
\end{array} \right).
\end{eqnarray}

It is obvious now that the lapse and shift functions enter the Hamiltonian
\begin{equation}
\label{dRGT}
H=-\int d^3 x\sqrt{\gamma}\left(N\left(\mathop{R}\limits^{({\mathit 3})}+\frac{1}{\gamma}\left(\frac12 \left(\pi^j_j\right)^2-\pi_{ik}\pi^{ik}\right)
-V\right)+2N^i\mathop{{\bigtriangledown}^k}\limits^{({\mathit 3})}\pi_{ik} \right)
\end{equation}
non-linearly in the $\sqrt{\gamma}NV$ term, and therefore, naively
one would expect to get some non-trivial equations for these unphysical variables instead of the physical constraints
${\mathcal C}$ and ${\mathcal C}^i$. If it was the case, we would end up with six degrees of freedom
including the Boulware-Deser ghost. However, the very peculiar feature of the potential (\ref{potential})
is that one combination of these constraints does survive as a physical relation leaving us with
a healthy number of degrees of freedom \cite{dRGT,dRGT2,dRGT3}. And we would like to understand the reasons
for that.

The standard approach to the 
Hamiltonian analysis \cite{HR2,HR3,HR4} is to explicitly calculate the square root matrix in
(\ref{potential}). Note that if it was not for the spatial metric $\gamma$, then the square root
would have been really easy to find. Indeed, it follows from the very simple relation:
\begin{eqnarray}
\label{property}
\left( \begin{array}{cc}
1 & a^i \\
-a^j & -a^i a^j 
\end{array} \right)^2=
\left(1-a^k a^k\right)
\left( \begin{array}{cc}
1 & a^i \\
-a^j & -a^i a^j 
\end{array} \right).
\end{eqnarray}
However, one can not just simply take the square root of this part and then combine it with the square root of
$\gamma^{-1}$ as they can never anticommute. But one can hope to redefine the shift functions
$N_i=(\delta_i^j+ND_i^j)n_j$ such that the remaining $N$-dependent part of the $\sqrt{g^{-1}f}$ would still
acquire this nice form \cite{HR2}:
\begin{eqnarray}
\label{decomposition}
\sqrt{g^{-1}f}=
\frac{1}{N\sqrt{1-n^kn^k}}
\left( \begin{array}{cc}
1 & n^i \\
-n^j & -n^i n^j 
\end{array} \right)+
\left( \begin{array}{cc}
0 & 0 \\
0 & X^{ij}(\gamma,n) 
\end{array} \right),
\end{eqnarray}
and then the anticommutator with the $X$ matrix would account for the difference between $N^i$ and $n^i$,
and between $X^2$ and $\gamma$.
If this is so, then after such a redefinition the lapse function will enter the Hamiltonian linearly
enforcing a truly physical constraint. Clearly, if there exists a combination of lapse and shifts which
enters the Hamiltonian linearly then it must be possible to perform such a decomposition of $\sqrt{g^{-1}f}$
after a linear in $N$ redefinition of shifts\footnote{If it was not linear then the lapse would definitely go non-linear
in front of ${\mathcal C}^i$´s.}. Indeed, in the (singular) limit of $N\to 0$, which is equivalent to
$n_i\to N_i$, the  $\frac1N$-part of the square root
must tend to satisfying the property (\ref{property}); and therefore it should always satisfy it because
the difference between $n_i$ and $N_i$ is determined by $N$ on which no explicit dependence is allowed.
And vice versa, if one finds such a change of variables, calculates
the $X^{ij}$ matrix and proves the relation (\ref{decomposition}), then the theory 
is free of the Boulware-Deser ghost. And it was actually done in \cite{HR2}
with the only limitation that the $D_i^j$ operator is determined as a non-linear function of $n$, and therefore, in
terms of initial variables, the transformation $D$ is found only as an implicit function of the lapse and shifts.

These results \cite{HR2} have proven that the dRGT gravity is a potentially healthy deformation of GR at the full
non-perturbative level. Moreover, it has been done for an arbitrary reference metric \cite{HR3} and for
non-minimal models too. The latter actually correspond to adding two more potential terms,
$V_2=\left({\rm Tr} \sqrt{g^{-1}f}\right)^2-{\rm Tr} \left(\sqrt{g^{-1}f}\right)^2$ and
$V_3=\left({\rm Tr} \sqrt{g^{-1}f}\right)^3-3\left({\rm Tr} \sqrt{g^{-1}f}\right)\left({\rm Tr} \left(\sqrt{g^{-1}f}\right)^2\right)
+2{\rm Tr} \left(\sqrt{g^{-1}f}\right)^3$.

\section{The simple argument}

In our approach we introduce an extra matrix of auxiliary fields $\Phi^{\mu}_{\nu}$ into the model,
so that the potential takes the following form:
\begin{equation}
\label{our}
V=\frac{m^2}{N}\left( \Phi^{\mu}_{\mu}+ \left(\Phi^{-1}\right)^{\mu}_{\nu}N^2 g^{\nu\alpha}f_{\alpha\mu}\right)
\end{equation}
which yields the standard $2m^2{\rm Tr}\sqrt{g^{-1}f}$ term (\ref{potential}) after integrating out the auxiliary
fields. With the Minkowski reference metric we can safely demand $\Phi_i^k=\Phi^i_k$ and $\Phi^0_i=-\Phi^i_0$, while in a
general case some other combinations of $\Phi$´s will drop out of the action.\footnote{If it does not seem so
obvious as for how to arrive at the form (\ref{our}), one may start with
$NV=2m^2\Phi^{\mu}_{\mu}+\kappa^{\mu}_{\nu}\left(\Phi^{\nu}_{\alpha}\Phi^{\alpha}_{\mu}-N^2 g^{\nu\alpha}f_{\alpha\mu}\right)$
and integrate out the $\kappa$´s.} Now, if we set out to make every single step explicitly, 
the primary constraints are ${\mathcal C}_1=\pi_N$, ${{\mathcal C}_2}_i=\pi_{N^i}$
and ${{\mathcal C}_3}^{\mu}_{\nu}=\pi_{\Phi^{\nu}_{\mu}}$. The constraint ${{\mathcal C}_3}$ will
generate the matrix constraint ${{\mathcal C}_4}=\Phi^2-N^2g^{-1}f$, and we are particularly interested
in the constraints ${{\mathcal C}_5}$ and ${{\mathcal C}_6}_i$ generated by ${\mathcal C}_1$ and ${{\mathcal C}_2}_i$
respectively.

With a simple commutation we obtain
$${{\mathcal C}_6}_i=\sqrt{\gamma}\left(-2\mathop{{\bigtriangledown}^k}\limits^{({\mathit 3})}\pi_{ik}+m^2 N^2 \left(\Phi^{-1}\right)^{\mu}_{\nu}
\frac{\partial}{\partial N^i}g^{\nu\alpha}f_{\alpha\mu}\right)$$
where the derivative of the matrix (\ref{basic}) can be easily found to be
\begin{eqnarray*}
N^2\frac{\partial}{\partial N^i}g^{\nu\alpha}f_{\alpha\mu}=
\left( \begin{array}{cc}
0 & \delta_i^j \\
-\delta_i^k & -\delta_i^j N^k- \delta_i^k N^j
\end{array} \right).
\end{eqnarray*}
And therefore these constraints allow us to determine the shift
functions in terms of $\gamma_{ik}$, $\pi^{ik}$ and $\Phi$; and a naive expectation would be that, in combination with
${\mathcal C}_4$, it is possible to express them solely in terms of $\gamma_{ik}$ and $\pi^{ik}$. 
Then we find the last remaining constraint at this stage
$${\mathcal C}_5={\mathcal C}^{(GR)}
+m^2\sqrt{\gamma} \left(\Phi^{-1}\right)^{\mu}_{\nu}
\frac{\partial}{\partial N}N^2g^{\nu\alpha}f_{\alpha\mu}
=\sqrt{\gamma}\left(-\mathop{R}\limits^{({\mathit 3})}-\frac{1}{\gamma}\left(\frac12 \left(\pi^j_j\right)^2-\pi_{ik}\pi^{ik}\right)
+2m^2 N\left(\Phi^{-1}\right)^i_j \gamma^{ij}\right)$$
where one should not worry too much about summing over two upper indices as we just
do not write out the unit matrix from the spatial part of the reference metric explicitly.\footnote{Note also 
that we could define $\Phi^2=g^{-1}f$ instead of $\Phi^2=N^2g^{-1}f$, and as one can easily show, it
would have resulted in the contribution to the constraint ${\mathcal C}_5$ which could be brought to a very
simple form, $2m^2\sqrt{\gamma}\left(\Phi^{-1}\right)^i_j \gamma^{ij}$, by use of the ${{\mathcal C}_4}=0$ equation.
This form contains neither the lapse nor shifts, although the lapse would have appeared in ${{\mathcal C}_6}_i$. 
And it would have been the first instance to face a crash of our naive expectations, that is, modulo some mixing, there are roughly three systems of
second class constraints: ${\mathcal C}_1$ and ${\mathcal C}_5$, ${\mathcal C}_2$ and ${\mathcal C}_6$, and 
${\mathcal C}_3$ and ${\mathcal C}_4$. However, despite its apparently striking form, this fact is not that easy
in being promoted to an actual proof, and therefore we will proceed with the definition (\ref{our}) which allows
for more pleasant calculations.}

Up to that stage of analysis, the total Hamiltonian density is
\begin{multline}
\label{ourham}
{\mathcal H}=-\sqrt{\gamma}N\left(\mathop{R}\limits^{({\mathit 3})}+\frac{1}{\gamma}\left(\frac12 \left(\pi^j_j\right)^2-\pi_{ik}\pi^{ik}\right)\right)-
2\sqrt{\gamma}N^i\mathop{{\bigtriangledown}^k}\limits^{({\mathit 3})}\pi_{ik}+
\sqrt{\gamma}m^2\left( \Phi^{\mu}_{\mu}+ \left(\Phi^{-1}\right)^{\mu}_{\nu}N^2 g^{\nu\alpha}f_{\alpha\mu}\right)+\\
+\lambda_1 \pi_N +{\lambda_2}^i\pi_{N^i}
+{\lambda_3}^{\mu}_{\nu}\pi_{{\Phi}^{\mu}_{\nu}}+
\sqrt{\gamma}{\lambda_4}^{\mu}_{\nu}\left(\Phi^{\nu}_{\alpha}\Phi^{\alpha}_{\mu}-N^2 g^{\nu\alpha}f_{\alpha\mu}\right)+\\
+\sqrt{\gamma}\lambda_5\left(-\mathop{R}\limits^{({\mathit 3})}-\frac{1}{\gamma}\left(\frac12 \left(\pi^j_j\right)^2-\pi_{ik}\pi^{ik}\right)+
2m^2 N\left(\Phi^{-1}\right)^i_j \gamma^{ij}\right)+\\
+\sqrt{\gamma}{\lambda_6}^i\left(-2\mathop{{\bigtriangledown}^k}\limits^{({\mathit 3})}\pi_{ik}+
2m^2\left(\left(\Phi^{-1}\right)^i_0+\left(\Phi^{-1}\right)^i_j N^j\right)\right),
\end{multline}
and it would be its final form, with the set of purely second class constraints, for a generic choice of potential. But in the case at hand
one can easily check that a particular
combination of unphysical momenta (see the subsection below)
does actually commute, in the weak sense, with the total Hamiltonian {\it irrespective} of the values
of Lagrange multipliers. Hence, the constraints ${\mathcal C}_4$, ${\mathcal C}_5$ 
and ${\mathcal C}_6$ do not allow to unambiguously express the naive unphysical variables, $N$, $N_i$ and $\Phi$,
in terms of the spatial metric and its momenta.
And we can nothing but conclude that they do contain a non-trivial constraining
equation for the would-be-physical variables $\gamma_{ij}$ and $\pi^{ij}$.\footnote{This is paralleled by the fact that
in the standard approach the total Hamiltonian commutes, at this stage of analysis, with $\pi_N$
with no restrictions on the values of Lagrange multipliers,
and indeed we have a physical constraint instead of an equation for $N$.} This observation
is actually the final step in our proof that the dRGT gravity contains strictly less than six degrees of freedom.

\subsection{Some technical details}

We now proceed to explicitly calculate the commutators of the unphysical momenta with the total Hamiltonian.
Obviously, within the constraint surface, we only need to commute them with the other constraints because
the commutations with the first line in the Hamiltonian (\ref{ourham}) have already been done, and the very
meaning of the other constraints is that those commutators do weakly vanish. Therefore, we can find
$$\frac{1}{\sqrt{\gamma}}\{ {\mathcal C}_1\ ,\ H\}=-2{\lambda_4}^i_k N\gamma^{ik}+2\lambda_5 m^2\left(\Phi^{-1}\right)^i_j \gamma^{ij},$$
$$\frac{1}{\sqrt{\gamma}}\{ {{\mathcal C}_2}_i\ ,\ H\}=
-2{\lambda_4}^i_0+2{\lambda_4}^i_kN^k+2m^2{\lambda_6}^k\left(\Phi^{-1}\right)^i_k,$$
$$\frac{1}{\sqrt{\gamma}}\{ {{\mathcal C}_3}^{\mu}_{\nu}\ ,\ H\}=2{\lambda_4}^{\mu}_{\alpha}\Phi^{\alpha}_{\nu}
+2m^2\lambda_5\gamma^{ij}N\frac{\partial}{\partial \Phi^{\nu}_{\mu}}\left(\Phi^{-1}\right)^i_j+
2m^2{\lambda_6}^i\frac{\partial}{\partial \Phi^{\nu}_{\mu}}\left(\left(\Phi^{-1}\right)^i_0+\left(\Phi^{-1}\right)^i_j N^j\right)$$
where we use the symmetry of the auxiliary fields and their Lagrange multipliers, 
${\lambda_4}_i^k={\lambda_4}^i_k$ and ${\lambda_4}^0_i=-{\lambda_4}^i_0$.
Normally, these commutators would give us fourteen independent linear equations for fourteen Lagrange multipliers,
so that all of them would be set to zero. However, using the ${\mathcal C}_4$ constraint and
a simple formula $\left(\Phi^{-1}\right)^{\alpha}_{\mu}\frac{\partial}{\partial \Phi^{\alpha}_{\nu}}\left(\Phi^{-1}\right)^{\beta}_{\kappa}=
-\left(\Phi^{-2}\right)^{\beta}_{\mu}\left(\Phi^{-1}\right)^{\nu}_{\kappa}$, one can deduce from 
the last commutator the following relations:
$${\lambda_4}^i_j=\frac{1}{2\sqrt{\gamma}} \left(\Phi^{-1}\right)^i_{\alpha}\{ {{\mathcal C}_3}^{\alpha}_j\ ,\ H\}+
m^2\left(\frac{\lambda_5}{N}\left(\Phi^{-1}\right)^i_j+\frac{{\lambda_6}^k\gamma_{kj}}{N^2}
\left(\left(\Phi^{-1}\right)^i_0+\left(\Phi^{-1}\right)^i_l N^l\right)\right),$$
$${\lambda_4}^k_0=\frac{1}{2\sqrt{\gamma}} \left(\Phi^{-1}\right)^k_{\alpha}\{ {{\mathcal C}_3}^{\alpha}_0\ ,\ H\}+
m^2\left(\frac{\lambda_5 N^j}{N}\left(\Phi^{-1}\right)^k_j+\frac{{\lambda_6}^i N_i}{N^2}
\left(\left(\Phi^{-1}\right)^k_0+\left(\Phi^{-1}\right)^k_l N^l\right)\right).$$
And we see that the combination of ${\lambda_4}^k_0-{\lambda_4}^k_j N^j$ depends
only on the commutators with ${\mathcal C}_3$, while both
$\lambda_5$ and $\lambda_6$ completely drop out of this expression. And therefore, using the commutator with ${\mathcal C}_2$,
we obtain
$$2m^2{\lambda_6}^k\left(\Phi^{-1}\right)^i_k=\frac{1}{\sqrt{\gamma}}
\left\{ {{\mathcal C}_2}_i+\left(\Phi^{-1}\right)^i_{\alpha}{{\mathcal C}_3}^{\alpha}_0-
\left(\Phi^{-1}\right)^i_{\alpha}N^j {{\mathcal C}_3}^{\alpha}_j\ , \ H\right\}$$
in the weak sense. (Note that one should not worry about commuting the Hamiltonian with the coefficients
in front of the constraints because the momenta do vanish on the constraint surface.) On the other hand,
one can compute ${\lambda_4}^i_j\gamma^{ij}$ and compare it with the $\{ {\mathcal C}_1\ ,\ H\}$ commutator. 
It (weakly) determines a linear combination of $\lambda_6$´s:
$$2m^2{\lambda_6}^k\left(\left(\Phi^{-1}\right)^k_0+\left(\Phi^{-1}\right)^k_l N^l\right)=
-\frac{1}{\sqrt{\gamma}}\left\{ N^2 \left(\Phi^{-1}\right)^i_{\alpha}\gamma^{ij}{{\mathcal C}_3}^{\alpha}_j+
N{\mathcal C}_1\ ,\ H\right\}.$$
These two results must agree, and it singles out a combination of momenta
$$\pi_N+N\left(\Phi^{-1}\right)^i_{\alpha}\gamma^{ik}\pi_{\Phi^k_{\alpha}}+
\frac{\left(\left(\Phi^{-1}\right)^i_0+\left(\Phi^{-1}\right)^i_l N^l\right)}{N}
\left(\left(\mathop{\Phi^{-1}}\limits^{({\mathit 3})}\right)^{-1}\right)_i^k
\left(\pi_{N^k}+\left(\Phi^{-1}\right)^k_{\alpha}\pi_{\Phi^0_{\alpha}}-\left(\Phi^{-1}\right)^k_{\alpha}N^j\pi_{\Phi^j_{\alpha}}\right)$$
which weakly commutes with the Hamiltonian for any values of the Lagrange multipliers.

In our approach, this combination determines the direction in the space of unphysical variables along which
there has been no restriction so far, under any of the constraints. This corresponds to the independence of $N$
in the standard treatment. In either approach, a constraint which commutes with
all the other constraints (and leaves one combination of the Lagrange multipliers undetermined) may appear in two
distinct situations: either it is a geniun first class constraint and corresponds to a gauge freedom in
the model, or some extra constraints are needed for the self-consistency so that the whole set
of constraints is non-degenerate second class. As the former seems not to be the case, the
generation of a one more constraint is unavoidable (and, at the end of the day, the values of all non-dynamical
fields should be somehow determined unless there is a gauge freedom indeed), and therefore the scenario forseen 
in the reference \cite{Kluson} is a priori highly implausible. 

Technically, what follows at the next step is that the Lagrange multipliers $\lambda_1$, $\lambda_2$, $\lambda_3$
should serve to ensure the preservation of ${\mathcal C}_4$, ${\mathcal C}_5$, ${\mathcal C}_6$ constraints. But
a factor in front of the all-commuting combination of momenta drops out of this game; and there would not be enough
freedom to make all the necessary commutators weakly vanish, if there is no gauge freedom. (And in our case the commutator
with the ${\mathcal C}_5$ constraint has two independent parts, proportional to the delta-function and to its derivative,
and the one undetermined combination of the other Lagrange multipliers would be used for fixing the latter part.) This is
how an extra constraint is generated in the model. However, it is not {\it a priori} clear that
it would be a physical one making the number of degrees of freedom not more than the healthy amount of
five (as opposed to five and a half) because the next constraint could just add the fourth equation for the lapse
and shift functions without any information on the spatial sector. Due to this reason it is harder to proof that the theory
contains five degrees of freedom than just to show that its number is less than six.
But a plausible argument would be that, after integrating out all but one undetermined unphysical variable, the remaining
combination should go linearly in the action with the coefficient equal to the remaining constraint which would therefore
commute with this part of the Hamiltonian (this is subject to the criticism in \cite{Kluson}). And as is 
known by now \cite{HR4}, the strange $5\frac12$-situation is not
the case, and {\it two} extra constraints do appear in the model, one is the "secondary" one for the spatial variables, and
the other finally fixes the lapse and shifts.

We would not proceed with explicit derivations in this paper because the most important result is already known \cite{HR2,HR4}, and 
our only purpose was to present an alternative and fairly simple method of analysis. However, all the necessary calculations are
very straightforward although time consuming.

\section{On arbitrary reference metrics\\ and non-minimal dRGT models}

Fow the sake of simplicity, up to now we have considered only the simplest choice of the reference metric,
i.e. the Minkowski one. However, the non-linear massive gravity has been proven to be free
of the Boulware-Deser ghost for any choice of the reference metric \cite{HR3}, and even in its bigravity
version too \cite{HR4}. Incorporation of an arbitrary lapse and an arbitrary spatial metric is actually trivial
(and the latter even makes the location of the spatial indices nicer), while shifts
do produce some problems and affect the simple form of the decomposition (\ref{decomposition}), see \cite{HR3}.
It can be readily seen by taking the general reference metric
\begin{eqnarray}
\label{ADMinv2}
f_{\mu\nu} = 
\left( \begin{array}{cc}
-\left( M^2-M_k M^k\right) & M_i \\
M_j & s_{ij}
\end{array} \right)
\end{eqnarray}
and calculating the basic building block of the model:
\begin{eqnarray}
\label{basic2}
g^{\mu\alpha}f_{\alpha\nu}=
\left( \begin{array}{cc}
\frac{M^2-M_k \left(M^k-N^k\right)}{N^2} & \frac{s_{ij}N^j-M_j}{N^2} \\
-\frac{N^j\left(M^2-M_k \left(M^k-N^k\right)\right)}{N^2}+\gamma^{ij}M_{j} & 
s_{ik}\gamma^{kj}-\frac{s_{ik}N^k N^j-N_i M^j}{N^2} 
\end{array} \right).
\end{eqnarray}

However, the complication is a relatively mild one: the
${\mathcal C}_5$ constraint receives a more involved contribution of
$2m^2N\sqrt{\gamma}\left(\left(\Phi^{-1}\right)^0_i M_j+\left(\Phi^{-1}\right)^k_i s_{kj}\right)\gamma^{ij}$ 
instead of the simple $2m^2N\sqrt{\gamma} \left(\Phi^{-1}\right)^i_j \gamma^{ij}$, and the form of ${\mathcal C}_6$ is also changed in an obvious way.
Nevertheless, it simply corresponds to a rotation of the variables,
and the subsequent calculations become a bit more complicated only due to the
form of the coefficients with no crucial change to the results. It actually should have been the case because, at least in the class of 
coordinate-independent reference metrics, a general metric can be transformed to 
zero shifts by a (linear) change of coordinates, e.g by the one which diagonalizes the matrix $f$.\footnote{Note 
that one {\it can}, of course, make a coordinate transformation, even in massive gravity, as long
as both the physical and the reference metric are being changed accordingly. And recall that the very property
of $\sqrt{g^{-1}f}$ to be decomposable into the sum of $N^{-1}$ and $N^0$ parts can be proven at each point of the
space-time manifold separately, with no reference to its coordinate dependence or independence.}

The models with general potentially ghost-free potentials can, in principle, be treated in the same way.\footnote{There is 
a somewhat subtle case of the purely quadratic potential $V_2$ for which the variation of the 
$\Phi^{\alpha}_{\alpha}\left(\Phi^{-1}\right)^{\mu}_{\nu}\left(g^{-1}f\right)^{\nu}_{\mu}-\left(g^{-1}f\right)^{\alpha}_{\alpha}$ term
with respect to $\Phi$ would leave an overall scalar factor undetermined much like the conformal invariance appears in the
Polyakov action of the bosonic string.}
Indeed, in order to check that the constraints ${\mathcal C}_1, \ldots, {\mathcal C}_6$ are preserved during the
evolution, one has to perform a straightforward computation of the commutators and to decide upon
the solvability of a system of linear equations for $\lambda_1, \ldots, \lambda_6$. And if some more constraints are required
then they definitely convey a non-trivial piece of information about the spatial sector of the model.

However, once we have understood the reasons for the sixth mode to be absent in
the minimal dRGT model, the ghost-free nature of the higher potentials can be better explained with
the standard argument of the references \cite{HR2,HR3}. We know that, after a linear in $N$ change of variables,
the minimal model action contains the lapse function only linearly.
And if one has convinced himself that it implies the decomposition (\ref{decomposition}), then it is also obvious \cite{HR2,HR3}
that the same is true of the $V_2$ and $V_3$ potentials because for a matrix $A$ with the property (\ref{property}) we have
${\rm Tr}(A^n)=({\rm Tr} A)^n$, and all unwanted powers of the lapse in the potential do cancel. This is how the symmetric
polynomials of the eigenvalues come into play; and in four dimensions there is only one more of 
them, ${\rm det}(g^{-1}f)$, which being multiplied by $\sqrt{-g}$ produces nothing but a constant shift of the action. 
Our method does not respect the decomposition (\ref{decomposition}), and therefore it is not so elegant in generalizing to
non-minimal models.

\section{Conclusions}

We have presented a new method of non-perturbative analysis of the non-linear massive gravity. And in particular,
we give a very simple argument for the absence of the Boulware-Deser ghost in this theory. The limitation
of our approach is that the argument is non-constructive, and does not proceed
directly in terms of the metric components. However, in principle, one can make all the
derivations explicitly, and calculate the number of constraints and independent degrees of freedom. In general, the power of this approach is
in the fact that many things can be done at the level of a bit bulky but absolutely straightforward calculations, 
very automatically, with no need of making more qualified and creative jobs such as taking the square root of a matrix.

Admittedly, the introduction of ten more configuration space dimensions (and ten more pairs of constraints to
eliminate them) is not very helpful for the actual physical calculations. But, given the relative ease with which
this argument shows the presence of extra constraints in the model, it is reasonable to hope that it can be
useful for discussing the formal aspects of the theory. And anyway, for the real calculations
a somewhat different language is better suited \cite{dRGT,dRGT2,dRGT3}. Obviously, it would be very interesting to find out
whether any kind of such formal tricks with auxiliary fields like $\Phi$ could produce a similarly simple argument
for the full non-linear stability in order-by-order perturbation theory.

{\bf Acknowledgements.} The author is very grateful to the Departamento de Fisica Teorica y del Cosmos of
the University of Granada, Spain where the main part of this work has been done, with the special thanks
to M. Bastero-Gil, M. Karciauskas and other members of the department for their hospitality. It is also a pleasure
to thank the ICTP and the organizers of the recent Workshop on Infrared Modifications of
Gravity, and the speakers at this very nice conference whose wonderful lectures comprised an invaluable
introduction to the subject.

\end{document}